\documentclass[prb,twocolumn,superscriptaddress,showpacs,amsmath,amssymb]{revtex4-2}
\usepackage{graphicx}
\usepackage{amsmath}
\usepackage{amssymb}
\usepackage{float}
\usepackage{color}

\usepackage{cancel,xcolor}
\usepackage{indentfirst}
\usepackage{CJKulem}
\usepackage{soul,xcolor}
\setstcolor{red}
\usepackage{ulem}
\usepackage{cancel}
\usepackage[pdfpagemode=UseNone,pdfstartview=FitH,colorlinks=true,linkcolor=blue,urlcolor=blue,anchorcolor=blue,citecolor=blue]{hyperref}

\def\sig{{\mbox{\boldmath{$\sigma$}}}}

\begin{document}

\def\sig{{\mbox{\boldmath{$\sigma$}}}}
\title {Nonlinear Hall Effect in Altermagnetic Warped Topological Insulators}
	\author{Debashree Chowdhury}
	\email{debashreephys@gmail.com}
		\affiliation{Centre for Nanotechnology, Indian Institute of Technology Roorkee, Roorkee, Uttarakhand-247667}

\begin{abstract}
Low-dimensional Dirac systems offer a versatile platform for quantum-geometric non-linear transport. Hexagonal warping in topological insulator (TI) surface states generates rich momentum-space geometric textures. However, mirror and time-reversal symmetries strictly constrain both the Berry curvature and quantum metric dipoles to zero. Here, we show that proximity-coupling a hexagonally warped TI to a $d$-wave altermagnet lifts spatial mirror symmetry via the interplay of $C_{3v}$ warping and $d$-wave spin-splitting. This symmetry breaking enables finite Berry curvature dipole components, which, however, remain heavily suppressed near charge neutrality, establishing the intrinsic quantum metric dipole as the sole driver of the non-linear Hall response. This scattering-independent response vanishes at $\mu = 0$ and exhibits odd parity under chemical potential inversion. Importantly, rotating the altermagnetic orientation angle $\phi$ continuously tunes quantum metric dipole and induces a sign reversal, peaking near $\phi=0$ and vanishing at $\phi = \pi/4$. Our findings highlight TI–altermagnet interfaces as ideal candidates for gate and orientation-tunable quantum metric spintronics.
\end{abstract}
\maketitle

\section{Introduction}
\begin{figure}[t]
   \centering
   \includegraphics[width=0.8\linewidth]{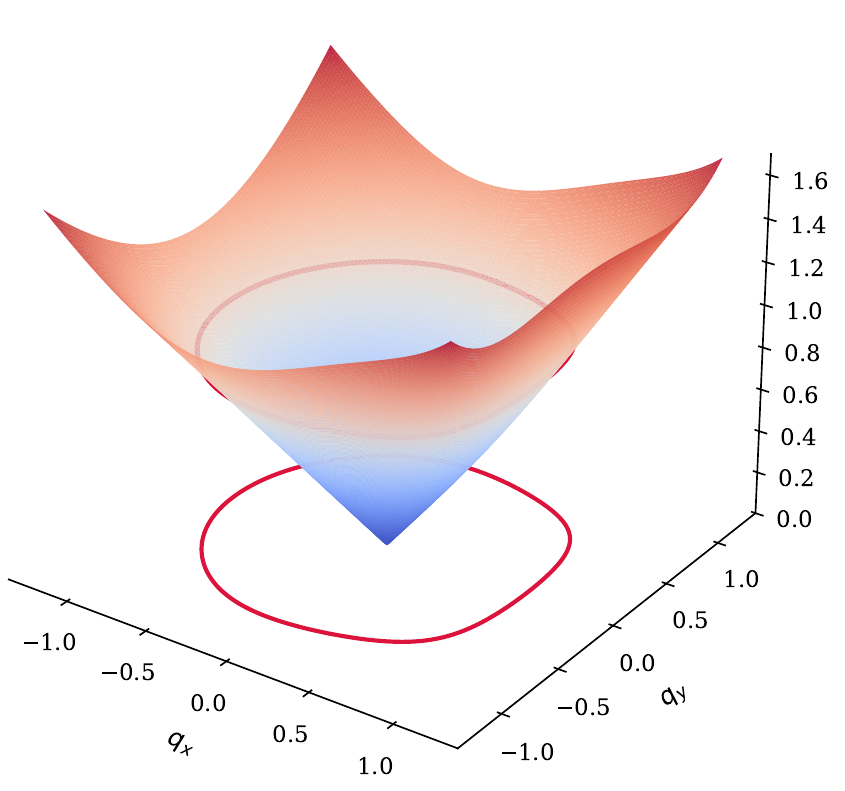}
   \caption{Schematic 3D energy dispersion $E_{+}(\mathbf{q})$ exhibiting anisotropic warping under $C_{3v}$ symmetry and $d$-wave spin splitting ($J_{\rm AM}$) with $\phi=0$. The broken mirror symmetry $M_y$ leads to an asymmetric Fermi surface contour (red line) at chemical potential $\mu$.}
   \label{fig1}
\end{figure}
The discovery of non-linear Hall effects has extended the scope of quantum geometric transport beyond linear response theory, uncovering rich physics in systems with broken inversion symmetry~\cite{Sodemann,Du1,Ortix, Bandyopadhyay,Ma2019}. Unlike the conventional linear Hall effect, which relies on net Berry curvature and external magnetic fields, second-order non-linear Hall conductivities describe a second-harmonic voltage response mediated by intrinsic and extrinsic quantum geometric properties in momentum space.

Initially, non-linear Hall transport was attributed primarily to the Berry curvature dipole \cite{Liao}, which measures the Fermi-surface momentum derivative of the Berry curvature distribution \cite{Sodemann, Du, Zhang}. However, recent theoretical advances have highlighted a distinct, purely intrinsic contribution mediated by the quantum metric, which is the real part of the quantum geometric tensor \cite{Gao, Wang, Ulrich, Verma, Gao2, Torma, Liu, Das, Gao2014, Sala2025}. The quantum metric dipole dictates intrinsic non-linear transport independently of scattering times \cite{Qiang, Yu,Tien}. A central challenge impeding the unambiguous experimental
demonstration of quantum metric transport is the prevalence of coexisting Berry curvature dipoles and extrinsic scattering backgrounds. In most
non-centrosymmetric or magnetic materials, the extrinsic and intrinsic components occur simultaneously. It is of great interest to explore systems where symmetry prevents one of those. 

In two-dimensional Dirac systems, such as the surface states of 3D topological insulators (TI), hexagonal warping introduces significant anisotropy into the linear Dirac dispersion~\cite{Fu}. While the warping induces rich momentum-dependent textures in both the Berry curvature $\Omega_z(\mathbf{q})$ and the quantum metric ${\cal G}_{ab}(\mathbf{q})$ \cite{Ezawa, Yar2022, Battilomo2019,He2021,Ezawa2024}, non-zero non-linear responses require the knowledge of fundamental symmetries. Under time-reversal symmetry ($\mathcal{T}$), the Berry curvature is odd under momentum reversal, whereas the quantum metric is even. Additionally, spatial mirror symmetry $M_y$ drives $(q_{x}, q_{y}) \to (q_{x}, -q_{y}),$ which strictly constrains second-order non-linear Hall transport. Generating a finite non-linear response requires breaking of either time-reversal or spatial mirror symmetry.

Altermagnetism ~\cite{Smejkal, Mazin,Feng2022}, which is a recently discovered magnetic phase characterized by collinear antiparallel order and $d$-wave (or higher-order) momentum-dependent spin splitting, provides an ideal platform to break spatial symmetries without generating net  magnetization. Proximity-coupling a hexagonally warped TI to a $d$-wave altermagnet \cite{AR} lifts spatial mirror symmetry $M_y$, establishing a competitive interplay among isotropic Dirac dispersion (see Fig.~\ref{fig1}), $C_{3v}$ warping, and $d$-wave spin splitting \cite{Fang2024}. This proximity effect requires a sufficiently thin interface barrier to induce a robust $d$-wave exchange field $J_{\rm AM}$ within the TI surface states. To preserve the non-trivial topology and maintain the validity of the effective $2\times 2$ surface Dirac Hamiltonian, $J_{\rm AM}$ must remain well below the bulk bandgap to avoid bulk-band hybridization, while simultaneously satisfying $J_{\rm AM} \gg k_B T$ to overcome thermal broadening and yield clear spin splitting.  Although nonlinear transport \cite{Gao} has been explored separately in warped TI \cite{Yar2022,Battilomo2019} under magnetic fields or strain, as well as in pure altermagnets \cite{Feng2022, Fang2024,Han,Chu2025}, the interplay between hexagonal warping and $d$-wave altermagnetism remains unexplored. 


In this work, we investigate the non-linear Hall effect arising from the interplay between hexagonal warping and $ d $ -wave altermagnetism from a quantum geometric perspective. A fundamental challenge in experimentally realising non-linear Hall transport is uniquely disentangling intrinsic geometric contributions from extrinsic scattering mechanisms, such as side-jump and skew scattering. TI– altermagnet heterostructures offer a remarkable platform for this purpose. Although the Berry curvature dipole components are heavily suppressed in the low-energy regime near charge neutrality, the intrinsic quantum metric dipole ($ D_ {\text {QM } } (\phi ) $ ) dominates the non-linear Hall response. Note, the orientation angle $ \phi $ controls the degree of spatial mirror symmetry breaking. Varying $ \phi $ allows direct tuning of the magnitude and sign of $ D_ {\text {QM } } (\phi ) $, with maximum transport response near $ \phi = 0 $ ($ d_ {x^ 2 - y^ 2 } $-wave symmetry). complete vanishing of $ D_ {\text {QM } } (\phi ) $ happens at $ \phi = \pi / 4 $ ($ d_ {xy } $-type splitting). Furthermore, this intrinsic quantum metric contribution is independent of the scattering time $ \tau $ and exhibits odd parity under chemical potential inversion ($ D_ {\text {QM } } (- \mu , \phi ) = -D_ {\text {QM } } (\mu , \phi ) $ ). The suppression of $ D_ {\text {BC } } $ near charge neutrality provides a clean, unobstructed regime to directly isolate and probe pure quantum metric effects solely via gate-voltage tuning and orientation alignment. We calculate $ D_ {\text {QM } } (\phi ) $ analytically and numerically, demonstrating that the interplay between $ C_ {3v } $ hexagonal warping and $ d $ -wave altermagnetic spin splitting generates finite, highly tunable non-linear Hall conductivities and characteristic non-monotonic resonance peaks as a function of Fermi energy. Our findings establish concrete guidelines for designing low-power, non-linear spintronic and optoelectronic devices based on TI– altermagnet heterostructures. \\


\section{Berry Curvature and Berry Curvature Dipole for the Altermagnet Model}
We start by adding a $d$ wave altermagnet to the hexagonally warped TI (see appendix A).
To account for an arbitrary orientation $\phi$ of the $d$-wave altermagnetic order parameter relative to the high-symmetry warping axes, the effective continuum Hamiltonian of the TI surface state is given by
\begin{align}
H(\mathbf{q}) = v_f (q_{x} \sigma_y - q_{y} \sigma_x) + d_z(\mathbf{q}) \sigma_z,
\end{align}
with the mass term defined as
\begin{align}
d_z(\mathbf{q}) &= \lambda (q_{x}^3 - 3q_{x} q_{y}^2) \nonumber\\&+ J_{{\rm AM}} \Big[ (q_{x}^2 - q_{y}^2)\cos 2\phi + 2 q_{x} q_{y} \sin 2\phi \Big].
\end{align}
Here, $\phi$ denotes the relative orientation angle between the principal $d$-wave altermagnetic axis and the crystal mirror plane. For $\phi = \pi/4$ ($d_{xy}$-type splitting), the $M_y$ mirror symmetry ($q_{y} \to -q_{y}$) is explicitly broken by the odd parity of the $2 J_{{\rm AM}} q_{x} q_{y} \sin 2\phi$ term.

The energy eigenvalues for the conduction ($+$) and valence ($-$) bands are
\begin{align}
E_{\pm}(\mathbf{q}) = \pm \Big\{ & v_f^2 (q_{x}^2 + q_{y}^2) + \Big[ \lambda (q_{x}^3 - 3q_{x} q_{y}^2) \nonumber \\
& + J_{{\rm AM}} \Big( (q_{x}^2 - q_{y}^2)\cos 2\phi + 2 q_{x} q_{y} \sin 2\phi \Big) \Big]^2 \Big\}^{\frac{1}{2}}.
\end{align}

The analytical expression for the Berry curvature is
\begin{widetext}
\begin{align}
\label{Berry}
\Omega_z^{\pm}(\mathbf{q}) = \frac{v_f^2 \Big[J_{{\rm AM}} \Big(q_{y}^2-q_{x}^2\Big) \cos (2 \phi )-2 q_{x} \Big(J_{{\rm AM}} q_{y} \sin (2 \phi )+\lambda  \Big(q_{x}^2-3 q_{y}^2\Big)\Big)\Big]}{2 \Big[\Big(2 J_{{\rm AM}} q_{x} q_{y} \sin (2 \phi )+J_{{\rm AM}} (q_{x}^2-q_{y}^{2}) \cos (2 \phi )+\lambda  q_{x} \Big(q_{x}^2-3 q_{y}^2\Big)\Big)^2+q_{x}^2 v_f^2+q_{y}^2 v_f^2\Big]^{3/2}}.
\end{align}
\end{widetext}
Following ref. \cite{Sodemann}, the Berry curvature dipole $D_{{\rm BC}}^a$ ($a \in \{x, y\}$) at zero temperature ($T = 0$) is given by
\begin{align}
\label{Da}
D_{{\rm BC}}^a = \oint_{{\rm FS}} \frac{dq_\parallel}{(2\pi)^2} \frac{v_a(\mathbf{q})}{|\mathbf{v}(\mathbf{q})|} \Omega_z(\mathbf{q}) \quad (a \in \{x, y\}),
\end{align}
where $v_a(\mathbf{q}) = \frac{1}{\hbar} \frac{\partial E_+}{\partial q_a}$ is the $a$-th Cartesian component of the band group velocity, and $|\mathbf{v}(\mathbf{q})| = \sqrt{v_{x}^2 + v_{y}^2}$ is its total magnitude. The ratio $\frac{v_a(\mathbf{q})}{|\mathbf{v}(\mathbf{q})|}$ represents the projection of the outward unit normal vector of the Fermi surface along the $a$-axis.

Due to the directional anisotropy arising from hexagonal warping and $d$-wave altermagnetism, the Fermi surface is non-circular (see fig. (\ref{fig1})). We numerically compute the integral in the polar coordinates $(q, \theta),$ with
$q_{x} = q \cos\theta$, ~~ $q_{y} = q \sin\theta$.
For a fixed chemical potential $\mu$, the Fermi momentum $q$ satisfies the implicit dispersion satisfies $E_+\big(q, \theta\big) = \mu$. Solving this we have
$dq_\parallel = \sqrt{q^2 + \Big(\frac{dq}{d\theta}\Big)^2} d\theta,$
where the implicit derivative is determined by $\frac{dq}{d\theta} = -\frac{\partial E_+ / \partial \theta}{\partial E_+ / \partial q}$. Substituting Eqs.~\eqref{Berry} into Eq.~\eqref{Da} yields
\begin{align}
\label{6}
D_{{\rm BC}}^a = \frac{1}{4\pi^2} \int_0^{2\pi} d\theta \, \frac{v_a(q, \theta)}{|\mathbf{v}(q, \theta)|} \, \Omega_z(q, \theta) \, \sqrt{q^2 + \Big(\frac{dq}{d\theta}\Big)^2} .
\end{align}
Numerical evaluation of this 1D contour integral reflects the stringent constraints imposed by crystal symmetries. Under spatial mirror operation $M_y (q_{x}, q_{y}) \to (q_{x}, -q_{y})$, the Berry curvature transforms as $\Omega_z(q_{x}, -q_{y}) = -\Omega_z(q_{x}, q_{y})$. For $\phi = 0$ ($d_{x^{2}-y^{2}}$-wave altermagnetism), the mass term $d_z(\mathbf{q})$ is strictly even under $M_y$, enforcing $D_{{\rm BC}}^y = 0.$ However $D_{{\rm BC}}^x$ remains symmetry-allowed. When the altermagnetic order parameter is rotated ($\phi \neq 0$), the $2 J_{{\rm AM}} q_{x} q_{y} \sin 2\phi$ term breaks $M_y$ mirror symmetry, allowing finite values for both Berry curvature dipole components $D_{{\rm BC}}^x$ and $D_{{\rm BC}}^y$. However, in the low-energy regime near charge neutrality, both components remain strongly suppressed compared to the quantum metric dipole ($D_{{\rm QM}}$), establishing $D_{{\rm QM}}$ as the primary driver of the observed non-linear Hall response.
\section{Quantum Metric and Quantum Metric Dipole}
\begin{figure*}[htbp]
  \centering
  \includegraphics[width=0.95\linewidth]{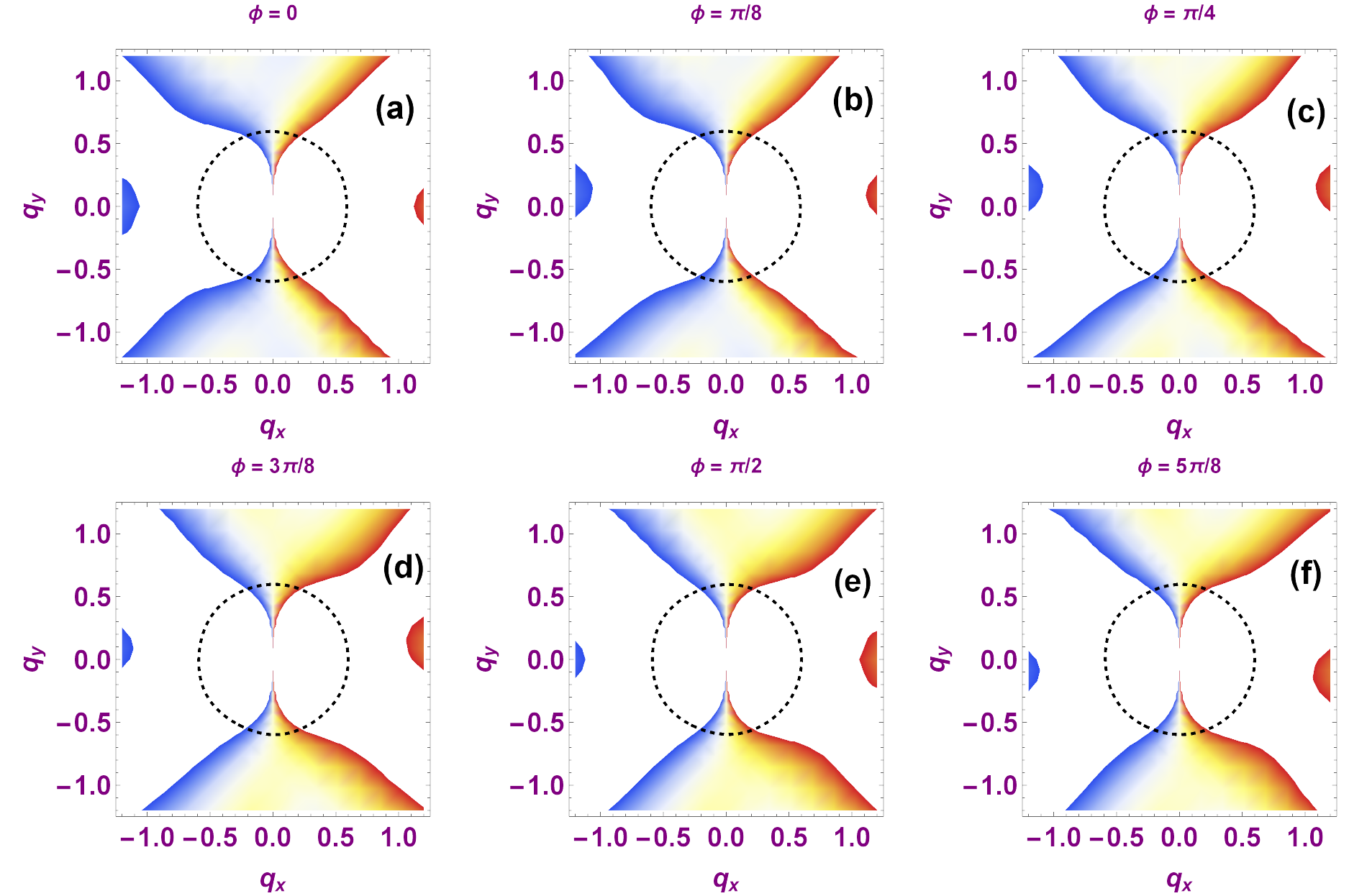}
  \caption{2D momentum-space distribution of the quantum metric dipole integrand $I(\mathbf{q}) = v_{x} {\cal G}_{yy} - v_{y} {\cal G}_{xy}$ with overlaid Fermi surface contours (dashed black lines) at chemical potential $\mu = 0.6$. The subpanels show orientation angles \textbf{(a)} $\phi = 0$, \textbf{(b)} $\phi = \pi/8$, \textbf{(c)} $\phi = \pi/4$, \textbf{(d)} $\phi = 3\pi/8$, \textbf{(e)} $\phi = \pi/2$, and \textbf{(f)} $\phi = 5\pi/8$. At $\phi = 0$, $M_y$ mirror symmetry breaking generates an asymmetric distribution of positive and negative hotspots, yielding a net positive $D_{{\rm QM}}$. As $\phi$ approaches $\pi/4$, diagonal mirror symmetry $M_{xy}$ renders the integrand strictly anti-symmetric across complementary quadrants, forcing $D_{{\rm QM}} = 0$. For $\phi > \pi/4$, the relative sign of the geometric hotspots inverts, reaching a maximum negative dipole response at $\phi = \pi/2$.}
  \label{fig2}
\end{figure*}

The quantum metric dipole tensor governs the intrinsic non-linear Hall response driven by the quantum geometry of the Bloch band. In two dimensions, the quantum metric tensor ${\cal G}_{ab}(\mathbf{q})$ is defined via the normalized field vector $\hat{\mathbf{d}}(\mathbf{q}) = \mathbf{d}(\mathbf{q})/E(\mathbf{q})$ as
\begin{align}
{\cal G}_{ab}(\mathbf{q}) = \frac{1}{4} \partial_a \hat{\mathbf{d}} \cdot \partial_b \hat{\mathbf{d}} = \frac{E^2(\partial_a \mathbf{d} \cdot \partial_b \mathbf{d}) - (\mathbf{d} \cdot \partial_a \mathbf{d})(\mathbf{d} \cdot \partial_b \mathbf{d})}{4 E^4},
\end{align}
where $a, b \in \{x, y\}$ and $\mathbf{d}(\mathbf{q}) = (-v_f q_{y}, v_f q_{x}, d_z(\mathbf{q}))$. In Appendix A and Appendix B, we show that without adding an altermagnet, the hexagonally warped TI surface states do not exhibit a non-linear Hall effect. Adding an arbitrary altermagnetic orientation $\phi$, the non-zero metric tensor components ${\cal G}_{yy}(\mathbf{q})$ and ${\cal G}_{xy}(\mathbf{q})$ take the compact form as 
\begin{widetext}
\begin{align}
&{\cal G}_{yy}(\mathbf{q}) = \frac{v_f^2}{4E^4} \Big\{ q_{x}^2 \Big[ v_f^2 + \lambda^2 (q_{x}^4 + 42 q_{x}^2 q_{y}^2 + 9 q_{y}^4) - 24 J_{{\rm AM}} \lambda q_{x}^2 q_{y} \sin(2\phi) + 4 J_{{\rm AM}}^2 q_{x}^2 \sin^2(2\phi) \nonumber\\&- 4 J_{{\rm AM}}^2 q_{x} q_{y} \sin(4\phi) \Big] + J_{{\rm AM}}^2 (q_{x}^4 + 6 q_{x}^2 q_{y}^2 + q_{y}^4) \cos^2(2\phi) + 2 J_{{\rm AM}} \lambda q_{x} (q_{x}^4 + 16 q_{x}^2 q_{y}^2 + 3 q_{y}^4) \cos(2\phi) \Big\}, \nonumber\\
&{\cal G}_{xy}(\mathbf{q}) =\frac{v_f^2}{4E^4} \Big\{ J_{{\rm AM}} \Big[ (q_{x}^2 + q_{y}^2) \Big( 3\lambda q_{y} (q_{y}^2 - 5q_{x}^2) \cos(2\phi) - 2 J_{{\rm AM}} q_{x} q_{y} \cos(4\phi) + 2\lambda q_{x} (2q_{x}^2 - 3q_{y}^2) \sin(2\phi) \Big)\nonumber\\& + J_{{\rm AM}} (q_{x}^4 - q_{y}^4) \sin(4\phi) \Big] - q_{x} q_{y} \Big[ v_f^2 + (q_{x}^2 + q_{y}^2) \Big( 2 J_{{\rm AM}}^2 + 3\lambda^2 (5q_{x}^2 - 3q_{y}^2) \Big) \Big] \Big\}.
\end{align}
\end{widetext}

To understand how a finite quantum metric dipole arises, we analyze the spatial mirror symmetry transformation $M_y (q_{x}, q_{y}) \to (q_{x}, -q_{y})$. Under $M_y$, the mass term $d_z(\mathbf{q})$ transforms as
\begin{align}\label{dz}
d_z(q_{x}, -q_{y}) &= \lambda (q_{x}^3 - 3q_{x} q_{y}^2) \nonumber\\&+ J_{{\rm AM}} \Big[ (q_{x}^2 - q_{y}^2)\cos 2\phi - 2 q_{x} q_{y} \sin 2\phi \Big].
\end{align}
For $\phi = 0$, the mass term $d_z(\mathbf{q})$ is strictly even under $M_y$. As a consequence, the velocity component $v_{x}(\mathbf{q})$ and metric component ${\cal G}_{yy}(\mathbf{q})$ are both even functions of $q_{y}$. The resulting dipole kernel $I(\mathbf{q}) = v_{x} {\cal G}_{yy} - v_{y} {\cal G}_{xy}$ remains symmetric under $q_{y} \to -q_{y}$, yielding a finite, non-vanishing quantum metric dipole $D_{{\rm QM}} \neq 0$. 
\begin{figure}[t]
   \centering   \includegraphics[width=0.99\linewidth]{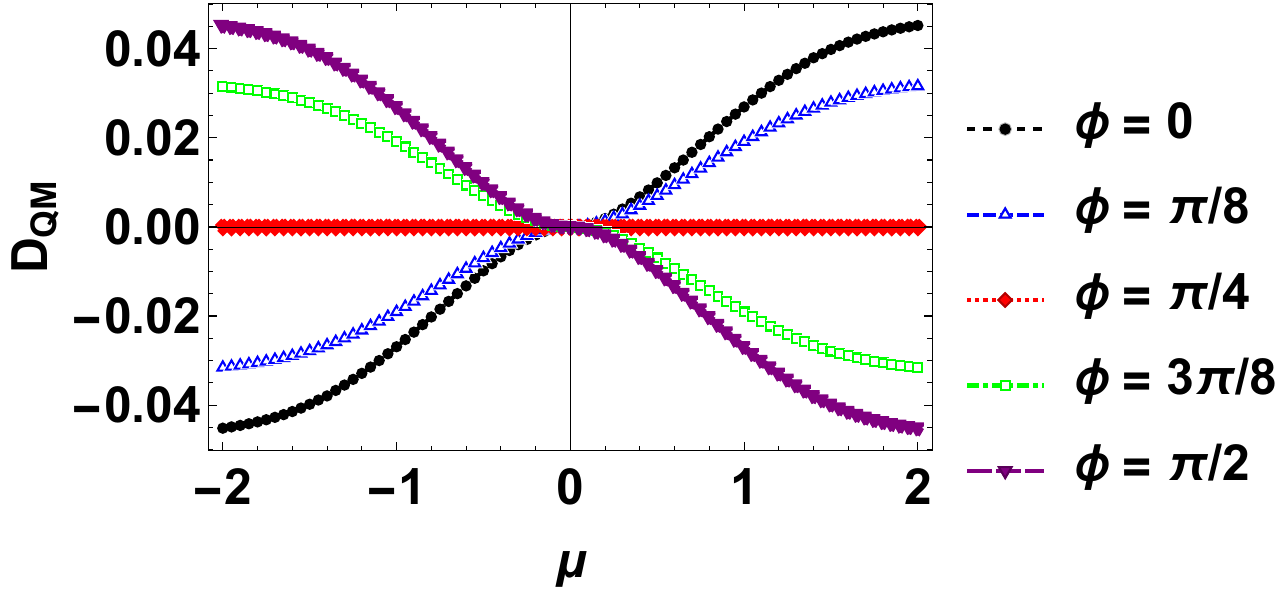}
    \caption{Quantum metric dipole $D_{{\rm QM}}$ as a function of the chemical potential $\mu$ for various orientation angles $\phi$ of the $d$-wave altermagnetic order parameter ($\phi = 0, \pi/8, \pi/4, 3\pi/8, \pi/2$) is plotted. The quantum metric dipole vanishes identically at the charge neutrality point ($\mu = 0$) and exhibits odd parity under chemical potential inversion, $D_{{\rm QM}}(-\mu) = -D_{{\rm QM}}(\mu)$. Due to spatial mirror symmetries, $D_{{\rm QM}}$ reaches its maximum magnitude near $\phi = 0$ ($d_{x^{2}-y^{2}}$-wave altermagnetism) and vanishes completely at $\phi = \pi/4$ ($d_{xy}$-type splitting).}
   \label{fig3}
\end{figure}
The momentum-space distribution of the quantum metric dipole integrand $I(\mathbf{q})$ and its evolution with the orientation angle $\phi$ are illustrated in Fig.~\ref{fig2}. At $\phi = 0$ ($d_{x^{2}-y^{2}}$-wave altermagnetism), the preservation of $M_y$ mirror symmetry in the mass term $d_z(\mathbf{q})$ generates a symmetric distribution of positive and negative geometric hotspots across $q_{y} \to -q_{y}$. The Fermi surface contour (dashed black curve) encloses a net positive excess in $I(\mathbf{q})$, giving rise to a maximal quantum metric dipole $D_{{\rm QM}} > 0$. As $\phi$ increases toward $\pi/8$, the geometric hotspots undergo a spatial rotation that reduces the net asymmetry across the Fermi surface. At the critical angle $\phi = \pi/4$ ($d_{xy}$-type splitting), the interaction between $C_{3v}$ hexagonal warping and the altermagnetic spin-splitting obeys diagonal mirror symmetry $M_{xy} (q_{x}, q_{y}) \to (q_{y}, q_{x})$. This renders the dipole integrand $I(\mathbf{q})$ strictly anti-symmetric across complementary quadrants, resulting in exact pairwise cancellation during the Fermi surface contour integration, thereby forcing $D_{{\rm QM}} = 0$. For orientation angles exceeding $\pi/4$, the spatial orientation of the positive and negative hotspots completely inverts. At $\phi = \pi/2$, the metric distribution becomes exact sign-inverted relative to $\phi = 0$, yielding a maximum negative dipole response $D_{{\rm QM}}(\pi/2) = -D_{{\rm QM}}(0)$.

This mirror symmetry breaking directly manifests in the quantum metric dipole kernel, $I(\mathbf{q})$. Because $I(q_{x}, -q_{y}) \neq -I(q_{x}, q_{y})$, the integrand loses its strict anti-symmetry, yielding a finite, non-vanishing contour integral over the Fermi surface
\begin{align}\label{DQM1}
D_{{\rm QM}} = \int \frac{d^2q}{(2\pi)^2}  I(\mathbf{q})\Big( -\frac{\partial f_0}{\partial E} \Big).
\end{align}
At zero temperature, $-\frac{\partial f_0}{\partial E}$ reduces to $\delta(E_+(\mathbf{q}) - \mu)$ confining the integral to the Fermi surface.

The numerical results for the quantum metric dipole $D_{{\rm QM}}$ as a function of the chemical potential $\mu$ across different orientation angles $\phi$ are summarized in Fig.~\ref{fig3}. As shown in Fig.~\ref{fig3}, $D_{{\rm QM}}$ vanishes identically at the charge neutrality point ($\mu = 0$) and exhibits strict odd parity under chemical potential inversion, $D_{{\rm QM}}(-\mu) = -D_{{\rm QM}}(\mu)$. Aligning with our mirror symmetry analysis, $D_{{\rm QM}}$ attains its maximum magnitude near $\phi = 0$ ($d_{x^{2}-y^{2}}$-wave altermagnetism) and is systematically suppressed as $\phi$ approaches $\pi/4$ ($d_{xy}$-type splitting), where it vanishes completely. Furthermore, $|D_{{\rm QM}}|$ displays a characteristic non-monotonic resonance peak at intermediate Fermi levels before decaying at high chemical potentials ($\mu \gg 0$), where the isotropic linear Dirac dispersion dominates over the hexagonal warping deformation.

The sign reversal of $D_{{\rm QM}}$ as the orientation angle crosses $\phi = \pi/4$ arises from the fundamental symmetry transformation of the $d$-wave altermagnetic spin-splitting relative to the $C_{3v}$ hexagonal warping. Rotating the altermagnetic order parameter past $\phi = \pi/4$ inverts the sign of the momentum-space mass term across complementary quadrants of the Fermi surface, effectively swapping the positive and negative contributions to the dipole integrand $I(\mathbf{q})$. At $\phi = \pi/2$, $D_{{\rm QM}}$ recovers its maximum magnitude but with a complete sign inversion relative to $\phi = 0$ (see Appendix C for detail). This sign flip is physically significant because it shows that the direction of the intrinsic nonlinear Hall current can be continuously toggled or inverted purely by reorienting the altermagnetic order parameter. Furthermore, it provides a definitive experimental signature for identifying the $d$-wave altermagnetic symmetry and mapping the underlying quantum geometric momentum distribution via direction-dependent nonlinear voltage measurements.

When driven by an AC electric field $\mathbf{E}(t) = \text{Re}\left[\mathbf{E}^{(\omega)} e^{i\omega t}\right]$, the second-harmonic current response is given by $J_a^{(2\omega)} = \chi_{abc} E_b^{(\omega)} E_c^{(\omega)}$, where $\chi_{abc}$ represents the non-linear conductivity tensor. Near charge neutrality, the Berry curvature dipole $D_{\text{BC}}$ is strongly suppressed, making the response predominantly governed by the intrinsic quantum metric contribution:
\begin{align}
\chi_{abc}^{\text{intrinsic}} = -\frac{e^3}{2\hbar^2} \epsilon_{ad} D_{\text{QM}, bc}^{d},
\end{align}
where $\epsilon_{ad}$ denotes the 2D Levi-Civita antisymmetric tensor ($\epsilon_{xy} = -\epsilon_{yx} = 1$) and $D_{\text{QM}, bc}^{d}$ is the quantum metric dipole. In contrast to extrinsic non-linear Hall effects arising from skew scattering or side jump, this intrinsic response is independent of the relaxation time $\tau$. Measuring the transverse second-harmonic signal $V_{2\omega} \propto \chi_{yxx}^{\text{intrinsic}}$ through standard lock-in detection therefore offers a scattering-free method to directly extract $D_{\text{QM}}(\phi)$ and observe its continuous sign reversal with the altermagnetic orientation angle $\phi$.

\section{Discussion}
The competitive interplay between $C_{3v}$ hexagonal warping ($\lambda$) and $d$-wave altermagnetism ($J_{{\rm AM}}$) provides a robust mechanism for inducing intrinsic non-linear Hall transport in two-dimensional Dirac systems. Note that neither hexagonal warping nor $d$-wave altermagnetic coupling alone breaks $M_y$ mirror symmetry to generate a net transverse response. Their joint interaction, however, breaks this spatial mirror symmetry, unlocking finite momentum-space Berry curvature dipole ($D_{{\rm BC}}$) and quantum metric dipole ($D_{{\rm QM}}$) contributions to the second-order non-linear Hall conductivity.

An interesting observation of our model is the continuous angular modulation and sign reversal of $D_{{\rm QM}}$ as a function of the orientation angle $\phi$. This is to be noted here that $D_{{\rm QM}}$ reaches its peak magnitude near $\phi = 0$ ($d_{x^{2}-y^{2}}$-wave phase) and it decreases toward $\phi = \pi/4$ ($d_{xy}$-type splitting). This is accompanied by a complete sign flip for $\phi > \pi/4$. In this situation, one recovers the maximum magnitude with inverted polarity at $\phi = \pi/2$. The fact that $D_{{\rm BC}}$ and $D_{{\rm QM}}$ exhibit distinct angular dependencies and parities under parameter tuning helps in measuring the directional sign flip in the second-harmonic transverse voltage upon rotating the drive current or magnetic order parameter. This provides a clear experimental fingerprint to identify the underlying $d$-wave altermagnetic order. Furthermore, this sign reversal enables a quantum geometric switch, where the net non-linear Hall current can be continuously tuned or toggled via spin-orbit torque or strain-induced domain reorientation.

Let us analyze the relevant physical parameters of the system in order to understand the balance between different extrinsic and intrinsic components.
Near $\mu \to 0$, we have $\hbar / \tau \ll |\mu|$. Considering Fermi velocity $v_f \sim 3 \times 10^5{\rm  m/s}$, hexagonal warping parameters $\lambda \sim 50-200{\rm  eV}\cdot{\rm \AA}^3$, and an altermagnetic spin-splitting $J_{{\rm AM}} \sim 10-50{\rm  meV}$, the intrinsic metric conductivity peaks sharply at small Fermi energies where the extrinsic skew scattering background is suppressed due to low density of states. For sample mobilities $\mu_{e} \sim 10^3-10^4{\rm  cm}^2/({\rm V}\cdot{\rm s})$ (corresponding to relaxation times $\tau \sim 0.1-0.5{\rm  ps}$), the intrinsic response dominates within a broad Fermi energy window around the resonance peak, $|\mu - \mu_{{\rm peak}}| \lesssim J_{{\rm AM}}$. However, the extrinsic side-jump and skew scattering contributions remain at least an order of magnitude smaller than $D_{{\rm QM}}$. Operating at low temperatures ($k_B T \ll \mu_{{\rm peak}} \ll T_N$) prevents thermal broadening of these quantum geometric features. Suitable platforms include warped 3D topological insulator thin films (${\rm Bi}_2{\rm Te}_3$, ${\rm Bi}_2{\rm Se}_3$) interfaced with insulating $d$-wave altermagnets (${\rm RuO}_2$, ${\rm MnF}_2$, ${\rm CrSb}$), where an AC drive field generates a measurable transverse second-harmonic current.

\section*{Acknowledgments}
D.C. acknowledges financial support from DST (project number DST/WISE-PDF/PM-40/2023).

\section*{Data Availability Statement}
The datasets generated and analyzed during the current study are available from the corresponding author upon reasonable request.
\appendix
\section{TI with Hexagonal Warping}

In the absence of altermagnetic coupling, the effective surface state Hamiltonian of a 3D TI with intrinsic hexagonal warping is given by~\cite{Fu}
\begin{align}
H(\mathbf{q})& = v_{f} (q_{x} \sigma_y - q_{y} \sigma_x) + \lambda (q_{x}^{3} - 3 q_{x} q_{y}^{2}) \sigma_z \nonumber\\&= \mathbf{d}(\mathbf{q}) \cdot \boldsymbol{\sigma},
\end{align}
where $\mathbf{q} = (q_{x}, q_{y})$, $v_{f}$ is the Fermi velocity, $\lambda$ is the hexagonal warping parameter, and $\boldsymbol{\sigma} = (\sigma_x, \sigma_y, \sigma_z)$ denotes the vector of Pauli matrices. The components of the vector field $\mathbf{d}(\mathbf{q}) $ are
\begin{align}
d_x(\mathbf{q}) &= -v_{f} q_{y},  d_y(\mathbf{q}) = v_{f} q_{x},\nonumber\\  d_z(\mathbf{q}) &= \lambda (q_{x}^{3} - 3 q_{x} q_{y}^{2}).
\end{align}

The corresponding energy eigenvalues for the conduction ($+$) and valence ($-$) bands are
\begin{align}
E_{\pm}(\mathbf{q}) =  \pm \sqrt{v_{f}^{2} (q_{x}^2 + q_{y}^{2}) + \lambda^2 (q_{x}^{3} - 3 q_{x} q_{y}^{2})^2}.
\end{align}

In polar coordinates ($q_{x} = q \cos\theta, q_{y} = q \sin\theta$), the dispersion reduces to
\begin{align}
E_{\pm}(q, \theta) = \pm \sqrt{v_{f}^{2} q^2 + \lambda^2 q^6 \cos^2(3\theta)},
\end{align}
which clearly illustrates the three-fold ($C_{3v}$) warping anisotropy.

\begin{figure}[t]
   \centering   \includegraphics[width=0.8\linewidth]{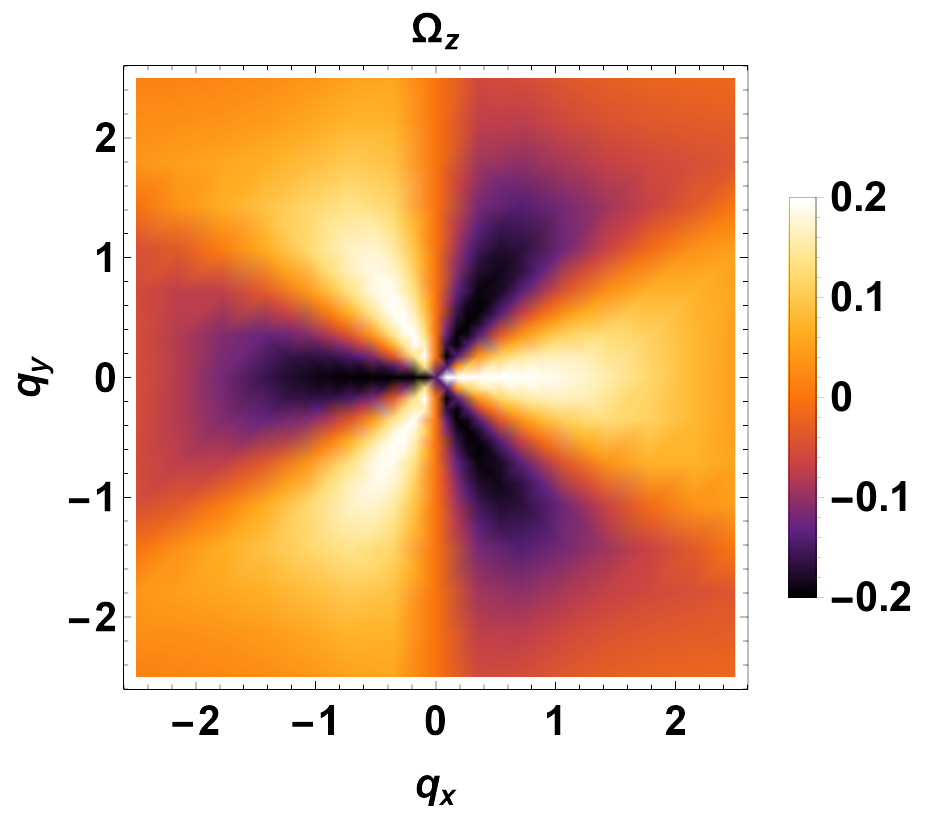}
    \caption{Berry curvature $\Omega_z(q_{x}, q_{y})$ is presented with $q_{x}$ and $q_{y}$ for a hexagonally warped TI surface state in the absence of altermagnetism. The distribution exhibits a clear six-fold pattern respecting spatial mirror symmetry $M_y$, which enforces exact cancellation across the Fermi surface and results in a vanishing net Berry curvature dipole.}
   \label{fig4}
\end{figure}
\subsection*{Berry Curvature and Net Dipole Integration}
The Berry curvature $\Omega_z(\mathbf{q})$ for the conduction band is expressed as
\begin{align}
\Omega_z(\mathbf{q}) = \frac{\lambda v_{f}^{2} q_{x} (q_{x}^2 - 3q_{y}^{2})}{\Big[ v_{f}^{2} (q_{x}^2 + q_{y}^{2}) + \lambda^2 (q_{x}^{3} - 3q_{x} q_{y}^{2})^2 \Big]^{3/2}}.
\end{align}
The momentum-resolved Berry curvature $\Omega_z(q_{x}, q_{y})$ for the unperturbed warped TI surface state is illustrated in Fig.\ref{fig4} . Due to the $C_{3v}$ symmetry of the hexagonal warping term ($\lambda$), the Berry curvature develops a signature six-fold sign-alternating texture in momentum space. Importantly, the distribution strictly preserves mirror symmetry $M_y$, obeying $\Omega_z(q_{x}, -q_{y}) = -\Omega_z(q_{x}, q_{y})$. Consequently, integration over the entire symmetric Fermi surface yields an exact cancellation, leaving the net Berry curvature dipole identically zero ($D_{\rm BC} = 0$) until time-reversal or mirror symmetries are broken by external couplings such as altermagnetism.
At zero temperature ($T = 0$), the Berry curvature dipole tensor $D_a$ ($a \in \{x, y\}$) over the Fermi sea is given by
\begin{align}
D_{BC}^{a} = \int \frac{d^2q}{(2\pi)^2} f_0(\mathbf{q}) \frac{\partial \Omega_z(\mathbf{q})}{\partial q_a} = -\int \frac{d^2q}{(2\pi)^2} \Omega_z(\mathbf{q}) \frac{\partial f_0(\mathbf{q})}{\partial q_a},
\end{align}
where $f_0(\mathbf{q})$ is the Fermi-Dirac distribution. Because time-reversal symmetry ($\mathcal{T}$) forces $\Omega_z(-\mathbf{q}) = -\Omega_z(\mathbf{q})$ and spatial mirror symmetry $M_y$ enforces $\Omega_z(q_{x}, -q_{y}) = -\Omega_z(q_{x}, q_{y})$, the net integration of $D_{BC}^{a} $ over the entire Fermi surface identically vanishes. 
\begin{figure*}[t]
   \centering
   \includegraphics[width=1.05\linewidth]{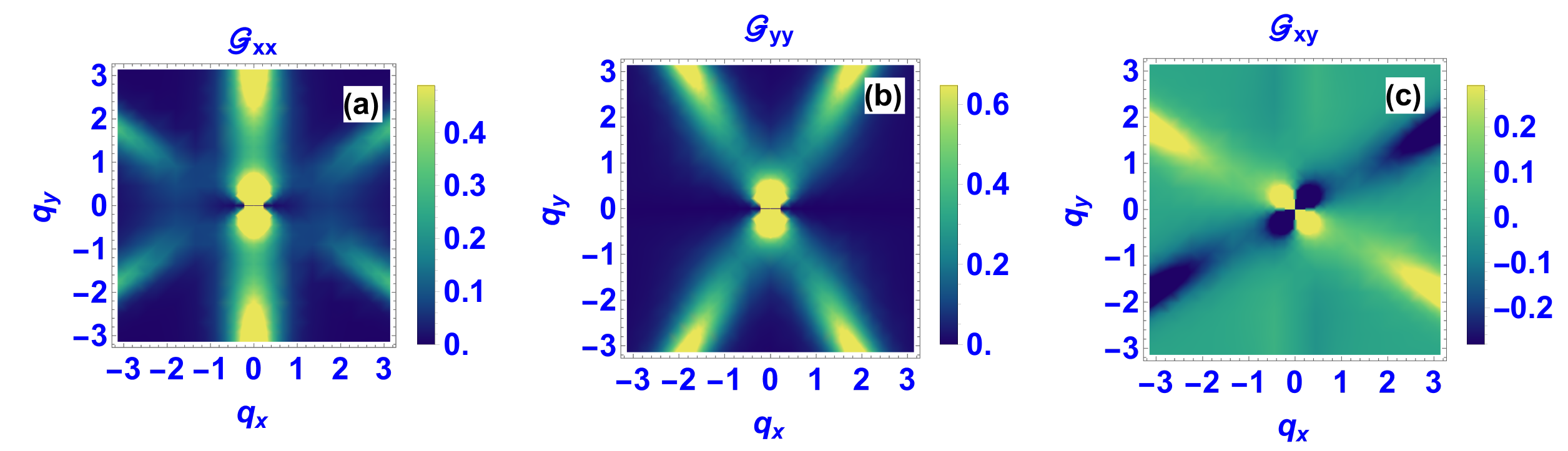}
    \caption{Momentum-space distribution of the quantum metric tensor components for a hexagonally warped TI surface state in the absence of altermagnetism (a) $\mathcal{G}_{xx}$, (b) $\mathcal{G}_{yy}$, and (c) $\mathcal{G}_{xy}$ plotted against $q_{x}$ and $q_{y}$. The preserved spatial mirror symmetry $M_y$ enforces $\mathcal{G}_{xx}(q_{x}, -q_{y})$ and $\mathcal{G}_{xy}(q_{x}, -q_{y})$ to transform as even and odd respectively, resulting in a vanishing net quantum metric dipole.}
   \label{fig5}
\end{figure*}
\subsection*{Quantum Metric Components}
The quantum metric tensor elements ${\cal G}_{ab}^{\pm}(\mathbf{q})$ ($a,b \in \{x,y\}$) describe the distance between neighboring quantum states in Hilbert space, and their momentum-space distributions for the unperturbed warped TI are shown in Fig.~\ref{fig5}.

Evaluating the components explicitly yields
\begin{align}
{\cal G}_{xx}(\mathbf{q}) &= \frac{v_{f}^{2}}{{\cal M}} \Big[ \Big(4 \lambda^2 q_{x}^6 - 18 \lambda^2 q_{x}^{4} q_{y}^{2}  \nonumber\\&+ q_{y}^{2} (9 \lambda^2 q_{x}^2 q_{y}^{4} + v_{f}^{2})\Big)+ 9 \lambda^2 q_{y}^6 \Big], \\
{\cal G}_{yy}(\mathbf{q}) &= \frac{q_{x}^2 v_{f}^{2} }{{\cal M}}\Big[ \lambda^2 (42 q_{x}^2 q_{y}^{2} + q_{x}^{4} + 9 q_{y}^{4}) + v_{f}^{2} \Big], \\
{\cal G}_{xy}(\mathbf{q}) &= \frac{q_{x} q_{y} v_{f}^{2}}{{\cal M}} \Big[ -3\lambda^2 (5 q_{x}^2 - 3 q_{y}^{2})(q_{x}^2 + q_{y}^{2}) - v_{f}^{2} \Big],
\end{align}
where ${\cal M}=4 \Big[ v_{f}^{2} (q_{x}^2 + q_{y}^{2}) + \lambda^2 (q_{x}^{3} - 3 q_{x} q_{y}^{2})^2 \Big]^2.$
The quantum metric dipole is written as
We choose  $\epsilon$ as the conduction band energy.
Before computing the integral, let us analyze the kernel of the integral $I(q))=[(v_{y}{\cal G}_{xx}-v_{x}{\cal G}_{xy})]$. The analytical expression is obtained as

\begin{align}
&I_{}(q_{x},q_{y})= \frac{q_{y} v_{f}^2 \Big(3 \lambda ^2 \Big(-6 q_{x}^2 q_{y}^{2}+7 q_{x}^{4}+3 q_{y}^{4}\Big)+v_{f}^2\Big)}{4 \Big(\lambda ^2 q_{x}^2 \Big(q_{x}^2-3 q_{y}^{2}\Big)^2+v_{f}^2 \Big(q_{x}^2+q_{y}^{2}\Big)\Big)^{3/2}}.
\end{align}
To calculate the quantum metric dipole, we compute the $\delta$ function in Eq. (\ref{DQM1}), and this could be done using the delta function property as follows
\begin{align}\label{A13}
\delta(g(x))=\sum_{i}\frac{\delta(x-x_{i}^{})}{|g'^{}_{}(x_{i})|}.
\end{align}
Importantly, as required by time-reversal symmetry (${\cal T}$), all quantum metric components remain even functions of momentum, $\mathcal{G}_{ab}(-\mathbf{q}) = \mathcal{G}_{ab}(\mathbf{q})$. Owing to the preserved spatial mirror symmetry $M_y$, the integrand of the quantum metric dipole, $I_{}(q_{x},q_{y})$, integrates to zero over the Fermi surface. Although, hexagonal warping induces rich local quantum metric distributions, a non-zero net quantum metric dipole $D_{\rm QM}$ cannot be generated unless $M_y$ mirror symmetry is explicitly broken, which is accomplished by adding the $d$-wave altermagnetic term $J_{\rm AM} (q_{x}^2 - q_{y}^{2}) \sigma_z$ in the main text.

\section{Symmetries for a Hexagonally Warped TI}
In this section we analyze the symmetry behavior of the momentum-space vector field, energy dispersion, group velocity, and quantum metric tensor under the spatial mirror reflection across the $y$-axis, $M_y$.
Under $M_y$, $q_{y}$ transforms to $-q_{y}$ and as a result, $d_z(q_{x}, -q_{y}) $ and 
$E(q_{x}, -q_{y})$ are ${\rm even}$ functions.
Consequently, the Fermi surface contour defined by the constraint $\delta(E(\mathbf{q}) - \mu)$ is strictly symmetric under $q_{y} \to -q_{y}$. The group velocity components $v_a(\mathbf{q}) = \frac{1}{\hbar} \frac{\partial E(\mathbf{q})}{\partial q_a}$ gives $v_{x}(q_{x}, -q_{y}) $ as ${\rm even}$ and 
$v_{y}(q_{x}, -q_{y}) $ as ${\rm odd}.$

The quantum metric components ${\cal G}_{ab}(\mathbf{q})$ depend on derivatives of the normalized vector field $\hat{\mathbf{d}}(\mathbf{q}) = \frac{1}{|\mathbf{d}|}(-v_{f} q_{y}, v_{f} q_{x}, d_z)$. The derivatives $
\partial_{q_{x}} d_z(q_{x}, -q_{y})$ and $
\partial_{q_{y}} d_z(q_{x}, -q_{y}) $ transform as even and odd under the mirror symmetry along $y.$

This eventually gives ${\cal G}_{xx}$ and ${\cal G}_{xy}$ are even and odd under $M_y$ respectively.
Combining these transformation rules into the quantum metric dipole kernel $I(q_{x}, q_{y}),$ one finds it is odd under $M_{y}.$ Since the Fermi surface integration domain is symmetric about $q_{y} = 0$, the integration in Eq. (\ref{DQM1}) yields zero identically.

\section{Analytical Derivation of Angular Dependence and Sign Reversal of $D_{{\rm QM}}(\phi)$}
The quantum metric dipole response is evaluated via the Fermi surface line integral, which is obtained as
\begin{align}
&D_{{\rm QM}}(\phi) = \frac{1}{(2\pi)^2} \int_{0}^{2\pi} \frac{q_{F}(\theta, \phi)}{|\mathbf{v}(q_{F}, \theta, \phi)|} \nonumber\\&\Big[ v_{x}(\theta, \phi) {\cal G}_{yy}(\theta, \phi) - v_{y}(\theta, \phi) {\cal G}_{xy}(\theta, \phi) \Big] d\theta,
\end{align}
where we write $q_{x} = q_{F} \cos\theta$ and $q_{y} = q_{F} \sin\theta$ in Eq. (\ref{DQM1}). To establish the analytical dependence of the quantum metric dipole $D_{{\rm QM}}$ on the orientation angle $\phi$, we recall the altermagnetic mass term in $d_{z}$ (see Eq. (\ref{dz})).
Under the diagonal spatial mirror operation $M_{xy} (q_{x}, q_{y}) \to (q_{y}, q_{x})$, the polar angle transforms as $\theta \to \pi/2 - \theta$. At the critical orientation angle $\phi = \pi/4$, the mass term satisfies
\begin{align}
&d_z(q_{y}, q_{x}, \pi/4) = \lambda (q_{y}^3 - 3 q_{y} q_{x}^2) + 2 J_{{\rm AM}} q_{y} q_{x} \nonumber\\&= -d_z(q_{x}, q_{y}, \pi/4) + 2 J_{{\rm AM}} q_{x} q_{y} - \lambda (q_{x}^3 - 3 q_{x} q_{y}^2),
\end{align}
where in the second line we change $\theta \to \pi/2 - \theta.$
This enforces a strict anti-symmetry in the quantum metric dipole kernel $I(\theta, \pi/4)$ across complementary quadrants of the Fermi surface
$I(\pi/2 - \theta, \pi/4) = -I(\theta, \pi/4).$
Integrating $I(\theta, \pi/4)$ over the symmetric $2\pi$ contour forces the net dipole response to vanish identically.

To capture the sign reversal behavior across $\phi = \pi/4$, we perform a Taylor expansion of the integrand around $\phi = \pi/4$ by defining the angular offset $\delta\phi = \phi - \pi/4$
\begin{align}
\cos(2\phi) &= \cos(\pi/2 + 2\delta\phi) \approx -2\delta\phi, \nonumber\\
\sin(2\phi) &= \sin(\pi/2 + 2\delta\phi)\approx 1 - 2(\delta\phi)^2.
\end{align}

Expanding the Fermi radius $q_{F}(\theta, \phi)$, group velocity $\mathbf{v}(\theta, \phi)$, and metric components ${\cal G}_{ab}(\theta, \phi)$ in powers of $\delta\phi$
\begin{align}
q_{F}(\theta, \phi) &= q_{F}^{(0)}(\theta) + \delta\phi \, q_{F}^{(1)}(\theta) + \mathcal{O}(\delta\phi^2), \nonumber\\
{\cal G}_{ab}(\theta, \phi) &= {\cal G}_{ab}^{(0)}(\theta) + \delta\phi \, {\cal G}_{ab}^{(1)}(\theta) + \mathcal{O}(\delta\phi^2).
\end{align}
Here, $q_F^{(0)}(\theta)$ and $\mathcal{G}_{ab}^{(0)}(\theta)$ denote the isotropic Fermi radius and quantum metric components evaluated at $\phi = \pi/4$, while $q_F^{(1)}(\theta)$ and $\mathcal{G}_{ab}^{(1)}(\theta)$ capture the first-order corrections in $\delta\phi$.
Substituting into the dipole integral yields
\begin{align}\label{DQMphi}
D_{{\rm QM}}(\phi) = D_{{\rm QM}}(\phi=\pi/4) + \delta\phi \Big. \frac{\partial D_{{\rm QM}}}{\partial \phi} \Big|_{\phi=\pi/4} + \mathcal{O}(\delta\phi^2).
\end{align}

The first term in Eq. (\ref{DQMphi}) is zero.  We next evaluate the derivative of the $D_{{\rm QM}}(\phi)$ as 
\begin{align}
\Big. \frac{\partial D_{{\rm QM}}}{\partial \phi} \Big|_{\phi=\pi/4} = \frac{1}{(2\pi)^2} \int_0^{2\pi} \mathcal{K}(\theta, \lambda, J_{{\rm AM}}, \mu) d\theta \equiv - C_0,
\end{align}
where $C_0$ represents the linear slope constant that governs how the quantum metric dipole changes sign as we rotate the altermagnetic orientation angle across $\phi.$
Thus, near $\phi = \pi/4$, the quantum metric dipole obeys a linear sign-changing relation
\begin{align}
D_{{\rm QM}}(\phi) \approx - C_0 \Big(\phi - \frac{\pi}{4}\Big).
\end{align}

This confirms that for $\phi < \pi/4$, $D_{{\rm QM}}(\phi) > 0$ and attends a positive maximum at $\phi = 0$. At $\phi = \pi/4$, $D_{{\rm QM}}(\pi/4) = 0$. For $\phi > \pi/4$, $D_{{\rm QM}}(\phi) < 0$, undergoing a complete sign inversion and reaching a negative maximum at $\phi = \pi/2$ with $D_{{\rm QM}}(\pi/2) = -D_{{\rm QM}}(0)$.


\end{document}